\documentclass[twocolumn,superscriptaddress,
	showpacs,amsfonts,amssymb,amsmath,floats,prb]{revtex4-2}
\usepackage{graphicx}

\usepackage[normalem]{ulem}

\usepackage{braket}

\def\e{\varepsilon}

\def\w{\omega}
\def\e{\epsilon}

\def\mat#1{\bm{#1}}
\usepackage{bm}

\begin{document}

\title{Spectral properties of strongly correlated multi impurity models in the Kondo insulator regime:
Emergent coherence, metallic surface states and quantum phase transitions}

\author{Fabian Eickhoff}
\author{Frithjof B. Anders}
\affiliation{Fakult\"at Physik, Technische Universit\"at Dortmund, 44221 Dortmund, Germany}

\date{July 27, 2021}

\begin{abstract}
We investigate the real-space spectral properties of strongly-correlated multi-impurity arrays
in the Kondo insulator regime. Employing a recently developed mapping onto an effective correlated cluster problem
makes the problem accessible to the numerical renormalization group. The evolution of the spectrum
as function of cluster size and cluster site is studied.  
We applied the extended Lieb-Mattis theorem to predict whether
the spectral function must vanish at the Fermi energy developing a true pseudo-gap or whether the spectral function remains finite at $\w=0$. Our numerical renormalization group spectra confirm the predictions of the theorem
and shows a metallic behavior at the surface of a cluster  prevailing in  arbitrary spatial dimensions.
We present a conventional minimal extension of  a particle-hole symmetric Anderson lattice model at $U=0$ that 
leads to a gapped bulk band but a surface band with mainly $f$-orbital character for weak and moderate hybridization strength. 
The change in the site-dependent spectra upon introducing a Kondo hole in the center of the cluster are presented as a function of the hole-orbital
energy. In particular the spectral signatures across the Kosterlitz-Thouless type quantum phase transition from a singlet
to a local moment fixed point are discussed.

\end{abstract}

\maketitle

\section{Introduction}

Multi-impurity Anderson models  (MIAM) \cite{Jones_et_al_1987,IJW92,Grewe91,EickhoffMIAM2020}
and its Kondo model counterparts have a long history in strongly correlated
materials. The basic properties of  Heavy Fermions (HF)  \cite{Grewe91,Maple95}
can be understood in terms of a hybridized band picture responsible for the 
heavy Fermi-liquid (FL)   \cite{Grewe1984-SSC} 
emerging from the coupling between the light itinerant conduction electrons and strongly localized and 
correlated $f$-orbitals. The discovery of superconductivity \cite{Steglich79} and competing magnetic order  \cite{Grewe91,LoehnysenWoelfeReview2007} in such materials has generated a large interest in this material class over  the last 40 years.

One special incarnation of such a system are Kondo insulators, where a correlation induced gap
opens up \cite{KASAYA1985,KasuyaKondoInsulator1996,SmB6trivalSurfaceConductor2018}: 
The resistivity increases below a characteristic temperature scale of the lattice, and the material crosses over from 
a metal to an insulator with a very small band gap when lowering the temperature.  Triggered by the advent of topological insulators,
such Kondo insulators have again drawn a lot of experimental and theoretical interest. Particular there is an active
debate on whether SmB$_6$ can be classified as topological Kondo insulator \cite{Kim2014,Park2016}
or a trivial surface conductor \cite{SmB6trivalSurfaceConductor2018}.  
Local density approximations \cite{Antonov2002} in combination with Gutzwiller methods
\cite{Lu2013} strongly indicate a topological nature of SmB$_6$, although the Kondo physics in such materials cannot
be addressed by those approaches. In topological insulators \cite{HasanKane2010} the  protection of the Chern number enforces metallic edge states.
This single-particle band properties, induced by spin-orbit coupling, inspired a lot of theoretical activity \cite{ColemanTopKI2010,Alexandrov2013,WernerAssaad2013,PetersKamakami2016} 
to address the experimental findings of a metallic surface state 
\cite{Kim2014,Park2016,SmB6trivalSurfaceConductor2018}
in Kondo insulators by converting the conventional Kondo lattice or periodic Anderson model (PAM) with a spin-orbit coupling term into a topological Kondo insulator in which a metallic surface emerges naturally from topological constrains. 

In this paper, we present site-dependent spectral functions of a MIAM
comprising a finite number of correlated orbitals embedded in a bi-partite lattice of uncorrelated orbitals.
Our conventional starting point \cite{Grewe1984-SSC,Kuramoto85} does not take
into account the spin-orbit coupling in the Ce 4f-shell considered by 
Dzero et al.\ \cite{ColemanTopKI2010} and, therefore, cannot address a topological Kondo insulator.

Raczkowski and Assaad reported a metallic surface layer in a finite size Kondo nano cluster 
studied with a finite temperature auxiliary-field quantum Monte Carlo (QMC) algorithm. The authors
realized that this  spectral feature  is unrelated to topological properties and interpret their results  \cite{RaczkowskiAssaad2019} as decoupling of the metallic surface from the bulk properties. 

We argue that the emerging metallic surface is a consequence of the 
extended Lieb-Mattis  theorem \cite{Shen1996,EickhoffKondoHole2021} 
that can be applied to arbitrary large finite size correlated cluster of a very simple MIAM.
In contrary, in a conventional pristine Kondo lattice model or periodic Anderson model the surface remains insulating 
at half filling. In this paper we present a simple argument based on this theorem how to predict whether
a local spectral function must vanish at $\w=0$ or a finite density of state is found. 
We  successfully employ this argument to understand the elementary  
real-space spectral properties of a MIAM cluster obtained with Wilson's numerical renormalization group (NRG)
approach \cite{Wilson75,BullaCostiPruschke2008}. In particular, the argument rigorously predicts a non-vanishing
spectral density in the $f$-orbitals spectra on the surface of a Kondo insulator cluster in agreement with 
auxiliary-field quantum Monte Carlo data \cite{RaczkowskiAssaad2019} and our NRG spectra.

In order to reconcile the insulating surface of a pristine Kondo insulator model 
with the findings of a metallic surface state in a finite size but arbitrary large Kondo insulator cluster embedded in a metallic host,
we address the question what is the minimum extension of a Kondo insulator to generate a metallic surface without
invoking the well discussed topological arguments. It turns out to be sufficient to include only one additional layer of non-interacting orbitals to generate a metallic surface band.  We present a mean-field tight-binding calculation
of a 2d Kondo lattice stripe to demonstrate the emerging of exactly two spin-degenerate metallic surface bands involving only the edge orbitals. 
While for weak hybridization, the surface band has mainly $f$-character close to Fermi energy, the uncorrelated edge orbital start to dominate once the hybridization strength exceeds the single particle hopping matrix element. 
This finding opens the door for speculations whether the metallic properties of the surface band in Kondo insulators might be
related to ruggedness of the surface edge.

We present the evolution of the spectral function in a finite size 1d MIAM 
to study the crossover from single-impurity surface physics to correlated bulk properties. We find that the gap size 
at the cluster center agrees  excellently with those obtained  from a  dynamical mean field (DMFT) calculation \cite{Georges96,KotliarVollhardt2004} for an infinitely large Kondo lattice model.
From the recently developed mapping of the problem onto an effective cluster \cite{EickhoffMIAM2020}
it becomes apparent that a single impurity Kondo model is insufficient to understand the screening of the local moments,
as already suspected by  Nozi\`eres \cite{Nozieres1985}. The spin moment screening mechanism 
in the Kondo lattice model can be better understood in terms of self-screening \cite{Grewe88} 
after mapping the model onto an emerging Hubbard model that is also responsible for the band formation. 

We extend our study to the site dependent spectral properties of the Kondo hole problem.
In particular, we track the spectral evolution close to the Kosterlitz-Thouless type
quantum phase transition (QPT) from a strong-coupling fixed point to a local moment fixed point describing
the local moment formation by a charge neutral substitution of a correlated orbital in a Kondo-insulator \cite{Schlottmann91I,EickhoffKondoHole2021}. The spectral properties explicitly reveals the occurrence of
a second Kondo temperature well below the lattice coherence temperature responsible for a sharp Kondo resonance that develops in the gap of the Kondo-insulator.

The paper is organized as follows:
Section \ref{sec:theory} is devoted to define the model under consideration and a very brief review
of our recently developed mapping onto an effective model \cite{EickhoffMIAM2020} which we employ to
calculate the spectral functions using the NRG. In particular, the matrix generalization of  the equation of motion
approach by Bulla et al \cite{BullaHewsonPruschke98} is presented. We present our results in Sec.\ \ref{sec:results}.
We start with the spectral evolution as function of the cluster size in Sec.\ \ref{sec:evolution-Nf}.
Sec.\ \ref{sec:lieb-mattis-theorem} is devoted to the application of the extended Lieb-Mattis theorem \cite{EickhoffKondoHole2021} to predict whether the spectral function must vanish at $\w=0$ or not. 
We demonstrate that our NRG calculations are fully in line with the prediction of this theorem,
providing a better understanding of the observed zero frequency features at certain lattice sites.
The site-dependent evolution of the spectral functions is presented in Sec.\ \ref{sec:concentrated-cluster-Nf7}
for the largest cluster we can access within the NRG. We augment our findings with
a tight-binding $U=0$ calculation for an infinitely large 2d stripe model with periodic boundary conditions in one dimension
to demonstrate the emergence of a metallic surface band in Sec.\ \ref{sec:metallic-surface}.
The last part of the result section, Sec.\ \ref{sec:Kondo-hole}, is devoted to the evolution
of the spectral properties in the vicinity of a Kondo hole. In particular, we trace the breakdown of the Kondo screening of
the emerging local moment close to the QPT. We close the paper with a short conclusion.

\section{Multi-impurity models and Kondo holes}
\label{sec:theory}

\subsection{Hamiltonian}
\begin{figure}[t]
\begin{center}
\includegraphics[width=0.45\textwidth]{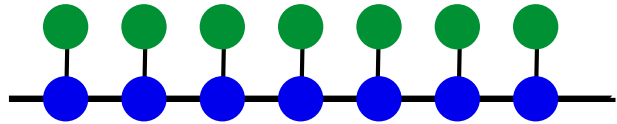}
\caption{Sketch of the geometric setup: An array of $N_f = 7$ correlated f-orbitals (green)  are locally 
coupled to a 1d tight binding chain (blue). 
}
\label{fig:1}
\end{center}
\end{figure}

The  MIAM \cite{EickhoffMIAM2020}
comprises of two subsystems:  the uncorrelated
conduction electrons and the localized correlated $f$-electrons.
The uncorrelated condition electrons are modeled by a tight-binding model
\begin{align}
 H_{\rm host}=\sum_{\substack{i,j, \sigma\\i\not=j}}\left(-t_{ij} c^\dagger_{i,\sigma}c_{j,\sigma}+\e^c_ic^\dagger_{i,\sigma}c_{i,\sigma}\right),
\label{eq:host}
\end{align}
where $t_{ij}$, $\e^c_i$ denote the transfer parameter and single particle energy, and $i,j$ labels the lattice sites of
the underlying lattice. We assume translational invariance of this subsystem so that this Hamiltonian is
diagonal in $k$ space and the single-particle dispersion is denoted by $\e_{\vec{k}\sigma}$. In principle a Zeemann energy
can be included but we only consider the system in the absence of an external magnetic field.

The localized $f$-electrons  are described the atomic part of a Hubbard Hamiltonian
\begin{eqnarray}
\label{eqn:himp}
H_{\text{corr}}& = & \sum_{l,\sigma}\epsilon^f_{l} f^{\dagger}_{l,\sigma}f_{l,\sigma}
+ H_{\rm int} \\
H_{\rm int} &=& \sum_{l} U_l f^{\dagger}_{l,\uparrow}f_{l, \uparrow}f^\dagger_{l,\downarrow}f_{l,\downarrow},
\end{eqnarray}
on a subset of $N_f$ lattice sites $l\in i$. $f_{l}^{(\dagger)}$ destroys (creates) an electron on impurity $l$, whose on-site energy is labeled by $\e^f_l$, and $U$ accounts for the on-site Coulomb repulsion.
A local single particle hopping term
\begin{align}
 H_{\rm hyb}=\sum_{l,\sigma}V_l c^\dagger_{l,\sigma}f_{l,\sigma}+\rm h.c.\,,
\label{eq:Hhyb}
\end{align}
facilitates the hopping of an electron between the localized $f$ orbital and the corresponding Wannier orbital
at the same lattice site. $V_l$ denotes the local hybridization of the impurity at lattice site $l$
with the corresponding local lattice orbital, such that the total Hamiltonian is given by
\begin{eqnarray}
H &=&  H_{\rm host} + H_{\text{corr}}  +  H_{\rm hyb}.
\label{eq:Hamiltonian}
\end{eqnarray}
For $l$  running over all lattice sites of the host, we obtain the periodic Anderson model.
If $N_f=1$, we recover the well understood single impurity model.
By setting individual values for $\e^f_l, U_l$ at specific lattice sites, we include the modeling of Kondo holes
into our rather general description \cite{EickhoffKondoHole2021}. Such a setup is schematically depicted
in Fig.\ \ref{fig:1} for $N_f=7$. We can also set $\e^f_l =\e_h^f$ at the location $h\in l$ of the Kondo hole, and sent $\e_h^f$ to very large value
to completely deplete that particular orbital.

\subsection{Mapping of the MIAM and Green's functions}

Here we briefly review the mapping of the MIAM 
onto an effective multi-band model in the wide band limit which we developed recently \cite{EickhoffMIAM2020}
and already applied to the Kondo-hole problem \cite{EickhoffKondoHole2021}.
In a translation invariant lattice the Hamiltonian $H_{\rm host}$
becomes diagonal in $\vec{k}$ space,
\begin{eqnarray}
H_{\rm host} &=& \sum_{\vec{k}\sigma} \e_{\vec{k}\sigma} c^\dagger_{\vec{k}\sigma} c_{\vec{k}\sigma}.
\end{eqnarray}
We define the coupling function matrix element $\Delta_{l,l'}(z)$ \cite{JabbenGreweSchmitt2012,EickhoffMIAM2020}
\begin{eqnarray}
\label{eq:hybridization-function}
\Delta_{l,l',\sigma}(z) &=& V_l V^*_{l'} D_{l,l'}^\sigma(z), 
\end{eqnarray}
where $D_{l,l'}^\sigma(z)$ denotes the unperturbed Green function of the non-interacting host in real-space,
\begin{eqnarray}
\label{eq:dll}
D^\sigma_{l,l'}(z) &=& \frac{1}{N} \sum_{\vec{k}} \frac{1}{z-\e_{\vec{k}\sigma}} 
e^{i\vec{k}(\vec{R}_l-\vec{R}_{l'})} .
\end{eqnarray}
Then, the exact real-space multi impurity Green's function (GF) is a matrix of dimension $N_f\times N_f$ and takes the form
\begin{eqnarray}
\label{eq:gf}
\mat{G}^f_{\sigma}(z) &=& [ z -\mat{E}_\sigma -\mat{\Delta}_\sigma (z) -\mat{\Sigma}_\sigma (z)]^{-1},
\end{eqnarray}
where the matrix $\mat{E}_\sigma$ contains the single-particle 
energies of the localized $f$-orbitals and $\mat{\Sigma}(z)$ denotes the self-energy matrix. Using the exact equation of motion for the GFs it is straight forward to show that
\begin{eqnarray}
\label{eq:sigma-eom}
\mat{\Sigma}_\sigma (z) &=& \mat{F}_\sigma[\mat{G}^f_{\sigma}] ^{-1},
\end{eqnarray}
and the matrix elements of the correlation matrix $\mat{F}$ are defined as \cite{BullaHewsonPruschke98}
\begin{eqnarray}
F_{l,l',\sigma} (z) = G_{B_{l\sigma}, f^\dagger_{l'\sigma}}(z), 
\end{eqnarray}
where the composite operator $B_{l\sigma}$ is generated by the commutator
\begin{eqnarray}
B_{l\sigma}(z) = [f_{l\sigma}, H_{\rm int}].
\end{eqnarray}

Since the host degrees of freedom are uncorrelated, the local host GF in the presence of the correlated orbitals 
obeys the exact relation
\begin{eqnarray}
\label{eqn:Gc}
G_{c_{i\sigma}, c^\dagger_{j\sigma}}(z)
&=& 
D_{i,j}^\sigma(z)
 \\ &&
+ \sum_{l_l,l_2}
D _{i,l_1}^\sigma(z) V_{l_1}  G^f_{l_1,l_2}(z)V^*_{l_2}  D_{l_2,j}^\sigma(z),
\nonumber
\end{eqnarray}
which can be casted into a matrix multiplication. While $D_{i,j}^\sigma(z)$ describes the free propagation of a conduction electron from the lattice site $j$ to the lattice site $i$, the t-matrix $V_{l_1}  G^f_{l_1,l_2}(z)V^*_{l_2}$
accounts for all the scattering processes of the conduction electrons on all the correlated impurity orbitals.

In the wide band limit, the influence of the host onto the dynamics of the correlated orbitals 
is determined by the real part and the imaginary part of $\mat{\Delta}_\sigma (z)$ at the Fermi-energy \cite{EickhoffMIAM2020}.

In a first step,
we diagonalize $\mat{\Gamma}_\sigma= \Im \mat{\Delta}_\sigma (-i0^+)$. The eigenvalues $\tilde V_l$ define
the new couplings to $N_f$ independent effective conduction bands \cite{EickhoffMIAM2020},
while the eigenvectors describe the composition of the new local orbitals in the rotated base. 

The real part $\mat{T}_\sigma= \Re \mat{\Delta}_\sigma (-i0^+)$ contains the host mediated  effective hopping matrix element  between the local orbitals, which are absorbed into the diagonal matrix
$\mat{E}_\sigma $ and rotated into the new local orbital basis. This procedure defines a mapping of the original
Hamiltonian, Eq.\ \eqref{eq:Hamiltonian}, onto an effective Hubbard model that couples to multiple effective conduction band channels.
Obviously, this Hubbard model cluster contains already the antiferromagnetic part of the RKKY interaction mediated by the conduction electrons.

Generically, a  multi-band model  with $N_f$ effective conduction electron channels 
is expected from the $N_f$ eigenvalues of $\mat{\Gamma}_\sigma$.
It turns out, however, that the encoded topology in $\mat{\Delta}_\sigma$
typically  leads to a reduced rank of $\mat{\Gamma}_\sigma$ \cite{EickhoffMIAM2020}. Since the upper bound of the rank is given by the number
of $\vec{k}$ vectors on the Fermi surface of the host, the problem reduces to an effective two-band model for arbitrary geometries in 1d, which makes the mapped MIAM fully accessible to the NRG.

Since the NRG always diagonalizes the rotated correlated orbital problem exactly in the first step, the complexity of the problem is essentially limited by the Hilbert space dimension of the correlated impurity problem. We are able to obtain very accurate NRG results up to $N_f=7$ correlated sites with the computer facilities we have access to.  We would like to refer the reader for more details on the approach and its limitation to our recent paper, Ref.\ \cite{EickhoffMIAM2020}.

The NRG applied to the mapped problem provides the correlation functions used to construct the
correlation self-energy matrix according to Eq.\ \eqref{eq:sigma-eom} after rotating back into the real-space orbital basis.
The full energy dependency of the uncorrelated host enters via Eq.\ \eqref{eq:gf} the $f$-orbital Green's function matrix.  
The equation of motion expression, Eq. \eqref{eqn:Gc}, provides the access to the real-space Green function matrix at
arbitrary lattice sites $i,j$ that can be chosen independently of the location of the correlated orbitals. Essentially the mapped MIAM serves as 
a powerful tool to calculate the real-space single particle Green function matrix including all spatial correlations of
the original MIAM.

\section{Results}
\label{sec:results}

\begin{figure}[t]
\begin{center}
\includegraphics[width=0.5\textwidth,clip]{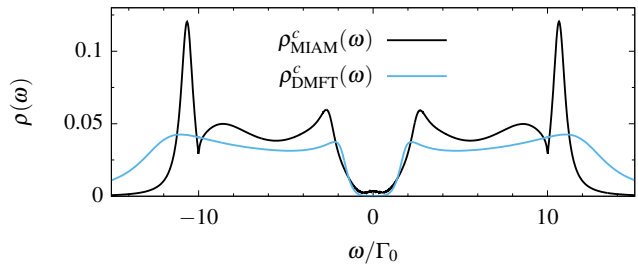}
\caption{$c$-spectral function for the central position in a $N_f=7$ MIAM where
$U=-2\e_f=10\Gamma_0$ in comparison with the DMFT solution of Kondo lattice model
with the identical Kondo coupling $g=\rho J=\rho J_{SW}=0.25$ obtained by the 
Schrieffer-Wolff transformation \cite{SchriefferWol66}.
}
\label{fig:2}
\end{center}
\end{figure}

In order to set the stage for our  detailed real-space investigation of the spectral functions
in a finite size MIAM, we plot a comparison between the $c$-orbital spectrum at the center, 
obtained via the exact equation of motion Eq.\ \eqref{eqn:Gc}, and the conduction electron
spectrum for a 1d Kondo lattice calculated within the dynamical mean field theory (DMFT)
\cite{Georges96} in Fig.\ \ref{fig:2}. 
We set the local Kondo interaction in such a way that the dimensionless Kondo coupling
$g=\rho J=\rho J_{SW}$ agrees with the prediction of the Schrieffer-Wolff transformation, $J_{SW}$ \cite{SchriefferWol66}
in the MIAM.

The data clearly show an excellent agreement between the DMFT and center spectra in the $N_f=7$
cluster. The cluster spectrum contains all many-body inter-site correlations exactly, which are only treated in mean-field within the DMFT.
In addition, the NRG broadening is overestimated in the effective site of the DMFT, hence the MIAM spectrum is much sharper
and contains sub-features at higher energies which are absent in the DMFT. In addition, the DMFT shows a clear pseudo-gap
with $\rho^c(\w=0)=0$ while we observe a finite value $\rho^c(\w=0)>0$ in the MIAM. This finite value can be explained within the extended
Lieb-Mattis theorem as demonstrated later on. In an infinitely large system, this finite value will disappear and the DMFT form will be asymptotically recovered.

\subsection{Evolution of the center spectral function  with the number of correlated sites $N_f$}
\label{sec:evolution-Nf}

We consider the geometry of Fig.\ \ref{fig:1} shown for $N_f=7$
and vary the number of correlated sites from $N_f=2$ to $N_f=7$.
The sites are labeled consecutively with  $l=1,\cdots, N_f$,
 $\epsilon_f=-U/2=5\Gamma_0$. 
The 1d conduction band host is particle-hole symmetric ($\e^c=0$), and the spectral functions are obtained
by an NRG calculation with the parameters stated in the caption and by employing the algorithm \cite{PetersPruschkeAnders2006,WeichselbaumDelft2007}  based  on the  complete NRG basis set
\cite{AndersSchiller2005,AndersSchiller2006}.

 \begin{figure}[t]
\begin{center}
\includegraphics[width=0.49\textwidth,clip]{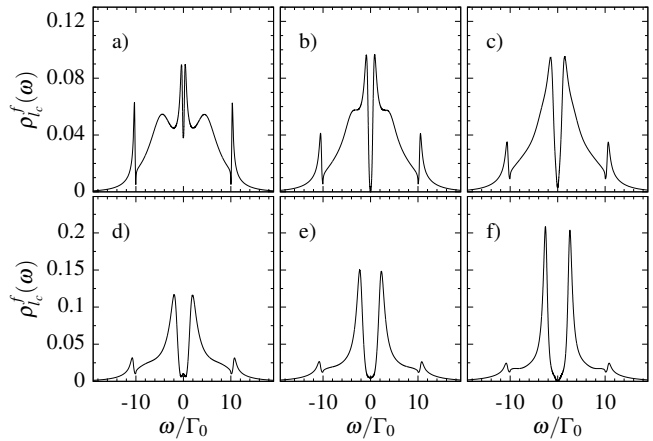}
\caption{Spectral function $\rho^f_{l_c}(\omega)$ of the central $f$-orbital in a dense array comprising (a) $N^f=2$, (b) $N_f=3$, (c) $N_f=4$, (d) $N_f=5$, (e) $N_f=6$ and (f) $N_f=7$ impurities. In case of even $N_f$ both central $f$-orbitals are equivalent due to parity symmetry. Parameters: $\epsilon_f=-U/2=5\Gamma_0$, 
$\e^c=0$,
$D=10\Gamma_0$, $\Lambda=2$ and $N_s=8000$.}
\label{fig:3}
\end{center}
\end{figure}

The evolution of the spectral function
at the center of the correlated sites, $l_c$, for $N_f=2-7$ respectively, is shown in Fig.\ \ref{fig:3}.
The local spectral function at site $l$ is extracted from the diagonal matrix element
\begin{eqnarray}
\rho_{l,\sigma}^f(\w) &=& \frac{1}{\pi} \Im G_{ll,\sigma}^f(\w -i0^+)
\end{eqnarray}
in the usual way. Note that we perform all calculations in the absence of an external magnetic field. Therefore, all
spectra in this paper are spin independent.

Let us first comment on the feature at $\w\approx D=10\Gamma_0$, visible in all spectra. This feature originates from the Van Hove singularity of the 1d density of states right at the band edge, which is included in $\mat{\Delta}(z)$, but doesn’t have any further physical meaning. Hence we will ignore these discontinuities in the following discussion.

A common feature is the formation of a gap structure symmetrically around $\w=0$, whose width is increasing with $N_f$.
This feature can be traced back to the spectral properties of a MIAM cluster by artificially setting $\mat{\Gamma}_\sigma=0$
and only include $\mat{T}_\sigma= \Re \mat{\Delta}_\sigma (-i0^+)$  in an exact diagonalization.   $\mat{\Gamma}_\sigma$ is responsible for the 
ferromagnetic component of the RKKY interaction \cite{EickhoffMIAM2020}, the remaining Kondo screening of a small cluster spin moment \cite{EickhoffMIAM2020}, renormalization of parameters and broadening of the spectrum.

We notice that $\rho^f_{l_c}(\w=0)$ at the center location remains finite and only vanishes for $N_f=3$ and $N_f=7$. 
Particularly, the finding of very small offset at $\w=0$ arises the question of whether this feature is caused by
a numerical problem or has a physical origin, since this has not been reported in 2d QMC spectra for a finite
size Kondo lattice cluster \cite{RaczkowskiAssaad2019}. We show in the next section that this has a physical origin,
and the extended Lieb-Mattis theorem can be used to predict under which circumstances
 $\rho^f_{l}(\w=0)=0$ and in which case $\rho^f_{l}(\w=0)$ must be finite at $T=0$.

For $N_f=2$, we have a realization of a two-impurity model where the energy dependent density of states break
the particle-hole symmetry in the even and the odd channels \cite{AffleckLudwigJones1995,LechtenbergEickhoffAnders2017,Eickhoff2018}, which generates 
the finite size hopping between the two correlated sites \cite{Eickhoff2018,EickhoffMIAM2020}. We obtain two Kondo peaks, one in the even and one in the odd channel, 
that are shifted away equally and in opposite direction 
from the chemical potential as a consequence of a non-resonant scattering. 
The symmetric superposition of the even and the odd spectral function yields 
the spectrum shown in Fig.\ \ref{fig:3}(a). This spectrum develops a pronounced valley between the two peaks but with a finite density of states at $\w=0$. The lower and upper charge excitations, sometimes referred to as Hubbard peaks, are well pronounced as known from the SIAM.

In Figure \ref{fig:3}(b) we proceed to the center orbital of the $N_f=3$ cluster. Here we have the first MIAM of second kind, where $N_f$ exceeds the number of screening channels \cite{EickhoffMIAM2020}. We have shown \cite{EickhoffMIAM2020} that the local moments of the two outer orbitals are aligned and AF correlated with 
the center spin. By inspecting  the spectrum in the vicinity of the chemical potential we find an exactly vanishing $\rho_{l_c}^f(\w=0)=0$.  Increasing the local cluster size to $N_f=4$ generates a small density of states, $\rho_{l_c}^f(\w=0)>0$. This observation holds up to $N_f=6$. The pseudo-gap width increases as well as  the peak height, and spectral weight is
transferred from the Hubbard peaks to the sharper features close to the Fermi level. The $f$-density of states resemble a band formation of the correlated orbitals by the effective hopping matrix $\mat{T}$, which asymptotically evolves in the heavy quasiparticles close to the Kondo insulator band gap \cite{PruschkeBullaJarrell2000}.
For a cluster size of 
$N_f=7$, the spectral function depicted in Fig.\ \ref{fig:3}(f) develops a true pseudo-gap with $\rho_{l_c}^f(\w=0)=0$.

\subsection{Applying the extended Lieb-Mattis theorem for predicting a pseudo-gap spectrum}
\label{sec:lieb-mattis-theorem}

We can combine an auxiliary Kondo problem with the extended Lieb-Mattis theorem
\cite{Shen1996,EickhoffKondoHole2021} to 
 decide whether the spectral function should have a finite value $\rho_{l}^f(0)$ at $\w=0$ or not. 
Shen  \cite{Shen1996} have proven
that for a number of $N_f$ local moments coupled by a local AF exchange interaction to a half-filled system of 
conduction electrons on a bipartite d-dimensional lattice with $N_c\geq N_f$ sites,  the ground state has a total $S_z$ component of
\begin{align}
S^\text{tot}_z = \frac{1}{2}|N_{c,\text{A}} - N_{c,\text{B}} + N_{f,\text{B}} - N_{f,\text{A}}|.
\label{eq:LiebMattis}
\end{align}
The extended version \cite{EickhoffKondoHole2021} generalizes Shen's version of the Lieb-Mattis theorem to  predict the ground state multiplet of a MIAM in the continuum limit, where the number $N_c$ of $c$-orbitals is considered to be infinitely large, $N_c\to\infty$.
Thereby, a finite size cluster is embedded into a conduction band continuum to account for a possible Kondo screening of
cluster ground state multiplets.

Setting aside subtle details of the spectrum close to $\w=0$,
the presented data in Fig. \ref{fig:3} suggest to qualitatively parametrize the density of states by a power law fit
$\rho_{l}^f(\w)\approx \rho_{l}^f(0) + C_{l}|\w|^r$  with an offset $\rho_{l}^f(0)$ and some exponent $r>1/2$.

Let us imagine, we extent the MIAM by an additional spin $1/2$ that is locally coupled antiferromagnetically with an arbitrarily  weak Kondo  coupling constant to the spin in the correlated 
orbital at site $l$. 
If the density of states is finite at $\w=0$, a Kondo screening of this
additional spin must occur in the limit $T\to 0$, and a strong coupling (SC) fixed point (FP) with a singlet ground state is found. 
If, however, $\rho_{l,\sigma}^f(0)=0$ and $r\geq 1/2$, we are dealing with
a pseudo-gap Kondo problem \cite{WithoffFradkin1990,Ingersent1996,FritzVojta2004} where the local
local moment (LM)  FP is stable in our parameter regime  for $T\to 0$.

Since the ground state of the total system remains independent of
the auxiliary AF coupling strength, we can now sent the auxiliary coupling  to infinity creating a local singlet from the auxiliary spin and the local moment at site $l$, such that the location of the fictitiously induced
local moment is shifted into the remaining system \cite{Clare96,Potthoff2014,EickhoffKondoHole2021}.

The essence of this first step is to map the problem of a finite spectral function $\rho_{l}^f(0)$ 
onto the question of a local moment formation in an effective MIAM where the correlated site $l$ has been removed. 
We extensively studied this so-called Kondo hole problem in a recent paper \cite{EickhoffKondoHole2021} where we have shown that the nature of the ground state can be derived from an
extended Lieb-Mattis  theorem on a bi-partite lattice with AF couplings at half filling.

In a second  step, we employ Shen's cluster Lieb-Mattis  theorem  \cite{Shen1996}  
for a finite size effective MIAM cluster \cite{Shen1996,Potthoff2014} with the $f$-orbital at site $l$  being removed, to determine its maximal spin component $S^\text{tot}_z$ of the ground state.
If $S^\text{tot}_z=0$, no effective moment remains and $\rho_{l}^f(\w=0)>0$. 
 
If, however $S^\text{tot}_z>0$, we have to be more careful and need to check in a third step 
whether by enlarging the conduction electron cluster, $S^\text{tot}_z$ is increasing or decreasing as outlined in Ref.\ \cite{EickhoffKondoHole2021}. This step extends finite size cluster to a true continuum model including additional screening channels due to the coupling to the host conduction electrons. If $S^\text{tot}_z$ increases, the effective exchange coupling 
of the ground state spin multiplet to the remaining conduction band channels is 
ferromagnetic, and a local moment prevails  leading to  $\rho_{l}^f(\w=0)=0$.
If, however $ S^\text{tot}_z$ decreases, we can conclude that the finite size cluster is AF coupled to the effective conduction band channel and a singlet ground state is found in the Kondo hole problem.

A specific example where this procedure agrees perfect with well known and established theoretical as well as experimental results is the case of graphene. Here the bi-partite lattice structure in combination with moderate correlations within the $\pi$-orbital enable the applicability of the Lieb-Mattis theorem, which predicts a stable local moment when a single carbon atom is removed. This is in agreement with theoretical studies on a microscopic level \cite{Yazyev2007,Palacios2008,Uchoa2008} and has already been observed experimentally \cite{Chen2014,AndreiGraphen2018,May2018}. At the same time, the pseudo-gap in the density of states of pristine graphene with an exponent of 
$r=1$  is well established as well, and the breakdown of the single ion Kondo effect 
was experimentally observed \cite{AndreiGraphen2018}.
Hence this example is in line with our argumentation that the occurrence of a stable local moment when a single site is removed is directly related to a local pseudo-gap spectrum with exponent $r\geq 1/2$.

Moreover, the predictions of the auxiliary model also fit to the $f$-orbital spectra of a half filled 2d PAM on simple cubic lattice with one $f$-orbital being removed, see Fig.\ 5 of Ref.\ \cite{Costa2018}.
According to the Lieb-Mattis theorem, removing another $f$-orbital on the same sublattice as the first one would increase the degeneracy of the ground state, whereas it is reduced when a $f$-orbital on the opposite sublattice gets removed. Consequently, the spectra of the $f$-orbitals on the opposite sublattice as the hole sites are expected to be finite at $\w=0$, while the others should be fully gapped.
Note that the precise value of $\rho^f(\w=0)>0$ can not be predicted by the Lieb-Mattis theorem. 
It is expected that $\rho^f(\w=0)>0$ falls of rapidly with increasing distance from the hole site, such that it might be difficult to resolve for larger distances as shown in Fig.\ 3 of Ref.\ \cite{Costa2018}.

We applied the extended Lieb-Mattis theorem to the situation where we removed the correlated orbital at the center position $l_c$ for the different cluster sizes $N_f$. Indeed, we only found a LM FP for $N_f=3$ and $N_f=7$ confirming our numerical findings as depicted in Fig.\ \ref{fig:3}.
For $N_f=2$, the auxiliary problem reduces to a SIAM that has a singlet ground state, and consequently $\rho_{1}^f(\w=0)>0$.

For $N_f=3$, the Kondo-hole cluster yield $2S^\text{tot}_z=1$, adding another conduction electron site yields $2S^\text{tot}_z=2$ indicating  a ferromagnetic coupling so that a the ground state carries a local moment. 
It is interesting to note that this $N_f=3-1=2$ effective Kondo-hole MIAM is identical
to a two-impurity Anderson model with both impurities located on the same sub-lattice. For this case we have explicitly proven in a NRG calculation including the full energy dependency of the conduction band host
that the LM fixed point is stable \cite{LechtenbergEickhoffAnders2017}. Therefore $\rho_{2}^f(\w=0)=0$, for $N_f=3$ as observed in Fig.\ \ref{fig:3}(b).

Note that the $N_f=3$ cluster is the minimal model to realize a MIAM of the second kind in a 1d host.
In a MIAM of the second kind the number of screening channels is less than the number of $f$-orbitals \cite{EickhoffMIAM2020}, which is required in order to realize a fully gapped spectra.
In Sec.\ III.1 of Ref.\ \cite{EickhoffKondoHole2021}, we presented a detailed analysis for the non-interacting limit of the simplest realization for a MIAM of the second kind in arbitrary dimensions on a simple cubic lattice, arranged as schematically depicted in Fig. 6 of Ref.\ \cite{EickhoffKondoHole2021} for the 1d and 2d case.
The even parity subspace of the model contains the central $f$-orbital ($f_{l_c}$) and the even linear combination of the outer ones ($f_+\propto \sum_{i\not=l_c}f_i$).
At low temperatures, only the center orbital, $f_{l_c}$, couples to the conduction band electrons, while the $f_+$-orbital does only indirectly couple via a finite hopping element $t_{{l_c}+}$ between the $f_{l_c}$- and $f_+$-orbital.
The Green's function of the $f_{l_c}$-orbital now involves propagations into the (decoupled) $f_+$-orbital and back, which, in case of PH symmetry, results in a diverging real part of the self-energy $\propto t^2_{{l_c}+}/\w$ and, consequently, in a fully gapped spectra - see Eq.\ (48) in Ref.\ \cite{EickhoffKondoHole2021}.
Since this contribution to the self-energy is still present when the correlations are switched on, the spectra in these models will always be fully gapped.
This argumentation also holds in 2d when $N_f=5$ $f$-orbitals are considered and arranged as depicted in Fig.\ 6 (b) of Ref.\ \cite{EickhoffKondoHole2021}.
In the 2d QMC study of Raczkowski and Assaad \cite{RaczkowskiAssaad2019}, however, no fully gapped spectrum was found for this specific $N_f=5$ model - see Fig.\ 3(a) in Ref.\ \cite{RaczkowskiAssaad2019}.
We  believe that the  mismatch is a finite  temperature effect, and encourage the authors of Ref.\ \cite{RaczkowskiAssaad2019} to adopt  the zero temperature project QMC  algorithm so as  to investigate in more details the long imaginary time behavior of the local $\tilde{f}$-Green  function.

For $N_f=4$, a removal of the $l=2$ f-orbital yields  $2S^\text{tot}_z=1$. By extending the cluster with one conduction electron site  \cite{EickhoffKondoHole2021} we can show that one of the two screening  channels is AF coupled and, therefore, the ground state of the extended Kondo-hole MIAM cluster has a singlet ground state. 
We leave it to the reader to apply the extended Lieb-Mattis theorem to show
that this prevails for $N_f=5,6$ while for $N_f=7$ a finite LM  FP 
is predicted by the Lieb-Mattis theorem which is compatible with our finding of 
$\rho_{4}^f(\w=0)=0$ for $N_f=7$.

The extended Lieb-Mattis  theorem has turned out to be a very powerful tool to rigorously establish the ground state degeneracy of auxiliary MIAM,  and allows us to make a prediction whether $\rho_{l}^f(\w=0)$ is finite or must vanish,
independent of the selected numerical approach for the calculation of the spectral functions.
This establishes the high accuracy of our NRG calculations presented in Fig.\ \ref{fig:3}, which are fully compatible
with the prediction of the extended Lieb-Mattis theorem.

\subsection{Concentrated MIAM cluster with  $N_f=7$}
\label{sec:concentrated-cluster-Nf7}

This section is devoted to the investigation of the real-space spectral functions
for a MIAM cluster depicted in Fig.\ \ref{fig:1}.

\subsubsection{Real-space spectral functions}

\begin{figure}[t]
\begin{center}
\includegraphics[width=0.49\textwidth,clip]{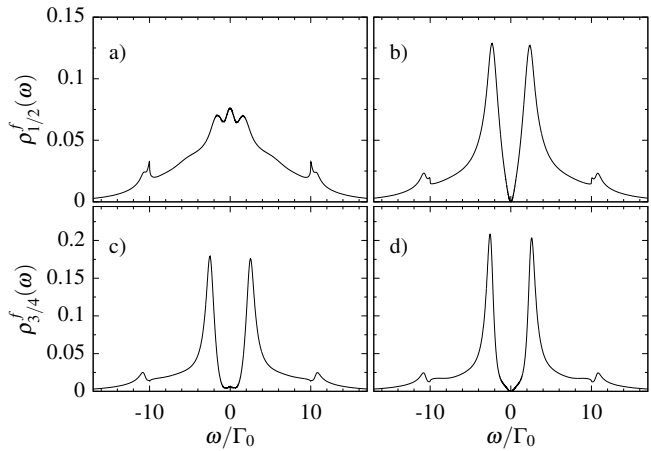}
\caption{Spectral functions $\rho^f(\omega)$ for an impurity array comprising $N_f=7$ $f$-orbitals. The panels depict the spectral function of the individual $f$-orbitals starting at the outermost in panel (a) and ending at the center in panel (d). Parameters: as in Fig. \ref{fig:3}}
\label{fig:4}
\end{center}
\end{figure}

After establishing the evolution of the spectral function at the central site with the cluster size $N_f$,
 we investigate the site depending spectra for our largest MIAM with $N_f=7$.
We present the $f$-spectra for the correlated site $l=1,2,3$ and the center location $l_c=4$ 
in Figs.\ \ref{fig:4}(a)-(d). Note that the data shown in Fig.\ \ref{fig:4}(d)
is identical to  Fig.\ \ref{fig:3}(f) and is added for completeness. 

A pseudo-gap feature is found in the spectra of all sites with two nearest neighbors.
This pseudo-gap  is absent on the surface of the 
correlated chain, i.\ e.\ at $l=1$ and $l=N_f$,  which resembles more a SIAM type spectral function. 
Since this feature has also been reported by Raczkowski and Assaad \cite{RaczkowskiAssaad2019} 
in an auxiliary QMC calculation for a 2d Kondo cluster model on a substrate, we believe this
is a generic surface property of a finite size MIAM in arbitrary spatial dimensions.

We can resort to the extended Lieb-Mattis theorem exemplified in our small clusters, 
that such metallic surface spectrum must indeed prevail in arbitrary
spatial dimensions and cluster sizes $N_f$, as long as the number $N_c$ of lattice sites of the underlying metallic host
is always significantly larger, $N_c >N_f$.  By removing an arbitrary correlated orbital on the 
surface of the cluster we are left with a new dense cluster of size $N_f-1$ which again
has a singlet ground state. The neglected coupling to the rest of the host does not change the ground state property. 
 The screening of an  auxiliary spin coupled antiferromagnetically to a surface $f$-orbital is 
always possible and, therefore, the density of states of the correlated orbital
at the surface for $\w=0$ must always remain finite. This general statement provides an
explanation for the finite $f$-spectral function $\rho^f(\w=0)$ at the surface of $6\times6$ Kondo cluster model \cite{RaczkowskiAssaad2019} and in our 1d correlated chain.

Let us proceed to the other correlated sites $l$ with $1<l<N_f$. At $l=2$ and $l_c=4$,  $\rho_{l}^f(\w=0)$ vanishes
exactly, while we find $\rho_{l}^f(\w=0)>0$
for $l=1,3$ in the data presented in Fig.\ \ref{fig:4}.
Again, this observation is consistent with the prediction of the extended Lieb-Mattis theorem. 
Since the number of uncorrelated sites are fixed and have a imbalance of $1$ for $N_f=7$,  
$S_z^{\rm tot}$ of the MIAM cluster
obtained from Eq.\ \eqref{eq:LiebMattis} is independent of the sublattice of the removed correlated orbital,
and we always find $S_z^{\rm tot}=1/2$. The cluster develops a local moment ground state
as expected from the isolated Kondo-hole problem \cite{EickhoffKondoHole2021}.  Now we add one more
uncorrelated orbital which must belong to the opposite sublattice as the first or last lattice site of the cluster.
While this leads to $N_{c,A}-N_{c,B}=0$ the total spin of the extended cluster depends on the imbalance 
$N_{f,B}-N_{f,A}$. Without loss of generality, let us assume that $l$ odd belongs to sublattice $A$. If we remove
a correlated f-orbital on sublattice $A$, we obtain $S_z^{\rm tot}=0$, and consequently a finite $\rho_{l}^f(\w=0)$,
while for even $l$ that belongs to sublattice $B$, the ground state spin is increasing to $S_z^{\rm tot}=1$.
Consequently, the effective MIAM cluster ground state spin is ferromagnetically coupled 
to the host for removing an $l$ even correlated orbital,  and the local moment cannot be fully screened. At these lattice sites, $\rho_{l}^f(\w=0)$ must vanish, confirming the findings of our NRG calculation.

\begin{figure}[t]
\begin{center}
\includegraphics[width=0.49\textwidth]{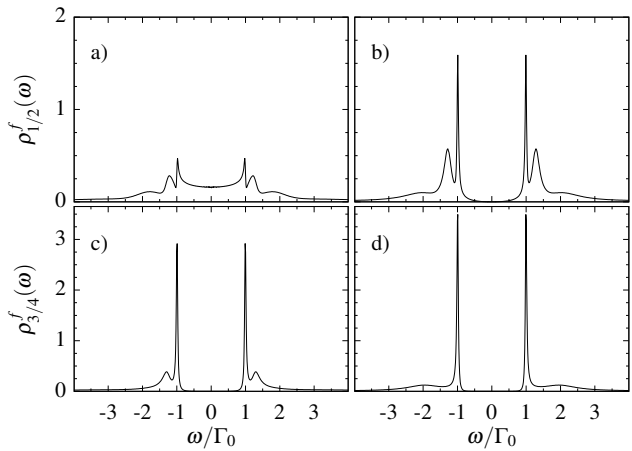}
\caption{Effective spectral functions $\tilde \rho^f(\omega)$ for an impurity array comprising $N_f=7$ uncorrelated
$f$-orbitals that are located at $\e_l^f=0$ to maintain half-filling. 
The panels depict the spectral function of the individual $f$-orbitals starting at the outermost in (a) and ending at the center in panel (d). 
}
\label{fig:5}
\end{center}
\end{figure}

The magnitude of $\rho_{l}^f(\w=0)>0$, however is not determined by the extended Lieb-Mattis theorem. 
In order to provide a basic understanding of the density of states, we assume that the MIAM cluster
shows Fermi liquid properties for $T\to 0$ implying that we can replace the full self-energy matrix $\mat{\Sigma}_\sigma (z)$ in Eq.\ \eqref{eq:gf} for $z\to 0$ by the Hartree contribution $\Re[\mat{\Sigma}_\sigma (0)]$. The diagonal matrix elements
renormalize the local single-particle $\e^f_l\to\e^f_l + U/2=0$ to ensure half-filling. The off-diagonal matrix elements 
renormalize the hopping matrix $\mat{T}$ but its magnitude is
analytically not known. 
Therefore, we calculated the spatially dependent spectral functions from
$\mat{\tilde G}^f_{\sigma}(z)=  [ z -\mat{\Delta}_\sigma (z) ]^{-1}$ in Fig.\  \ref{fig:5}
by neglecting the finite $U$ off-diagonal Hartree terms, essentially defining the $U=0$ and $\e^f_l=0$ problem.

Interestingly, the surface spectrum $\rho_{1,\sigma}^f(\w=0)$ 
remains metallic while all bulk spectral functions develop a finite gap. This gap corresponds exactly to the 
analytic gap of the 1d PAM located at $\pm\Gamma_0$. This confirms that a metallic surface and bulk 
gap-formation is already present in the $U=0$ PH symmetric MIAM.  In the corresponding  local conduction electron spectra
calculated by the exact equation of motion in Eq.\ \eqref{eqn:Gc} we found fully gapped spectra for all uncorrected conduction electrons
at the sites $l$ -- not shown here.

We still need to reconcile the even-odd 
oscillations in $\rho_{l}^f(\w=0)$ for $U>0$ with the finding of a gap for all bulk sites $1<l<N_f$ when $U=0$.
A more careful analysis presented in Ref.\ \cite{EickhoffKondoHole2021} relates  that 
to a $U=0$ Kosterlitz-Thouless type transition in the model where the f-orbital at site $l$ has been removed.
To elaborate on this difference  we still use the auxiliary spin picture for probing the local $f$-spectrum and
employ the $U=0$ supercell analysis for the MIAM presented in Ref.\ \cite{EickhoffKondoHole2021}.

In an isolated cluster of non-interacting orbitals and half-filling we have shown that removing a single  $f$-orbital at site $1<l<N_f$ generates a localized state. For a nearest neighbor tight-binding model, this localized orbital is located at the neighboring $f$-sites and at the conduction band Wannier orbital $l$ \cite{EickhoffKondoHole2021}. 
Therefore, the $f$-electron spectra must have a gap which we interpret as a precursor of the bulk gap.

Now we switch on a finite $U$. As we have demonstrated in Ref.\ \cite{EickhoffKondoHole2021},
the localized bound single particle orbitals can acquire a U-induced effective Kondo coupling such that the low-temperature local moments are screened at $T=0$. This provides the microscopic mechanism that is implicitly accounted for by the extended Lieb-Mattis theorem which is only valid for $U >0$ and half-filling. 
Since this effective Kondo coupling increases with some power law $U^\alpha$, $\alpha>1$  \cite{EickhoffKondoHole2021} for small $U$, this mechanism also explains that the zero frequency peaks for $U/\Gamma_0=20$ as shown in Fig.\ \ref{fig:7} are larger than those in Fig.\ \ref{fig:4} for $U/\Gamma_0=10$.
Eventually, for very large $U$, the crossover temperature $T_0$ at which the local moment is screened will decrease again as depicted in Fig.~15 of Ref.\ \cite{EickhoffKondoHole2021}.

Since the effective Kondo coupling for screening the local moment depends on its distance from the edge,
this mechanism is increasingly suppressed when moving from the edge to the center (bulk). Therefore, the pseudo-gap becomes more pronounce, and the finite $\rho_{l}^f(\w=0)$ is reduced beyond detectability for large cluster sizes $N_f\to \infty$,
even when the extended Lieb-Mattis theory predicts $\rho_{l}^f(\w=0)>0$.

Since the dimensionality of a cluster does not enter the theorem, we expect that such A-B sublattice oscillations in the local $f$-spectra  should also be present in a 2d Kondo cluster for $U>0$. Raczkowski and Assaad~\cite{RaczkowskiAssaad2019}, however, do not address this question, and the presented data in Fig.\ 3 of Ref.\ 
\cite{RaczkowskiAssaad2019} suggests a true pseudo-gap away from the cluster surface. 
This might be linked to the limits of the  employed auxiliary QMC.
In our NRG calculations we had to access exponential small temperature scales well below the single impurity Kondo temperature
to reveal these A-B sublattice oscillations in the f-spectra.

Also note that the extended Lieb-Mattis theorem cannot make a prediction of the magnitude of $\rho_{l}^f(\w=0)$ and only can help distinguishing between a finite $\rho_{l}^f(\w=0)$ and  $\rho_{l}^f(\w=0)=0$. 

Setting aside the even-odd oscillations of $\rho_{l}^f(0)$, we clearly can recognize the evolution of the spectral function
from an impurity type spectrum more towards a bulk  spectrum with a well pronounced pseudo-gap that is compatible in width with the DMFT solution of the 1d Kondo lattice problem.

\begin{figure}[t]
\begin{center}
\includegraphics[width=0.49\textwidth,clip]{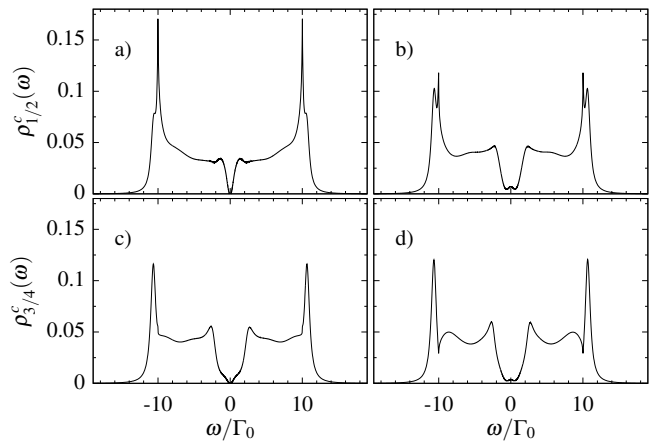}
\caption{The conduction electron spectral functions $\rho^c_l(\omega)$ 
for the same parameters as in Fig.\ \ref{fig:4}
(a) $l=1$, (b) $l=2$, (c) $l=3$, (d) $l=4$, 
The Green's functions was calculated via Eq.\ \eqref{eqn:Gc}.
}
\label{fig:6}
\end{center}
\end{figure}

In Figure \ref{fig:6} we present the spectra for the 
corresponding conduction electron orbital at the cluster site $l$. The Green's function was 
calculated  via Eq.\ \eqref{eqn:Gc}. The van-Hove singularities of the 1d simple cubic density of states of the host 
at the band edges are clearly visible. Additionally, a gaped density of states is also found at the surface of the cluster. This is for arbitrary $N_f$
and is in agreement with the findings in Ref.\ \cite{RaczkowskiAssaad2019}. Note that the data of Fig.\ \ref{fig:6}(d)
was already used for the comparison between the  DMFT Kondo lattice solution and the cluster calculations 
as presented in Fig.\ \ref{fig:2}.

\begin{figure}[t]
\begin{center}
\includegraphics[width=0.49\textwidth,clip]{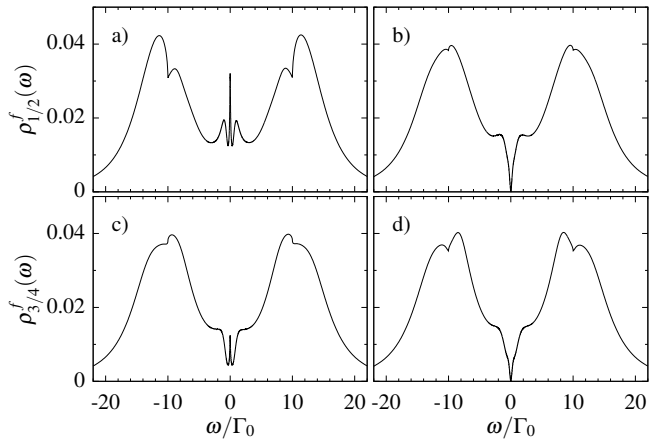}
\caption{Same as in Fig. \ref{fig:4} but with $\epsilon_f=-U/2=10\Gamma_0$.}
\label{fig:7}
\end{center}
\end{figure}

Reminiscence of the Hubbard side peaks expected at $\w\approx \pm U/2$
in the spectra can only be
seen for $l=1$ in Fig.\ \ref{fig:4}, while they disappeared for $1<l<N_f$. We attribute this
to the strongly energy dependent self-energy matrix $\mat{\Sigma}(z)$ as well as the peculiar feature of a 1d density of states
that is included in $\mat{\Delta}_\sigma (z)$ which shows to square root type divergencies at both band edges. 

We can, however, recover the Hubbard side peaks
by increasing $U$. The spectral functions for $U=20\Gamma_0$ are shown in Fig.\ \ref{fig:7} using the same NRG parameters as in Fig. \ref{fig:4}.  For each of the calculations we have used the adequate NRG iterations such that the spectrum remains unaltered when increasing the number of iteration to ensure that we are well in the low-temperature fixed point and $T\to 0$.

In addition, a low energy Kondo peak emerges on odd lattice sites. The hopping matrix 
$\mat{T}_\sigma$ generates an antiferromagnetic coupling of the different cluster local moments which leave a $S=1/2$ ground state multiplet. The bulk gap is already seen in the ED spectrum of the cluster. The remaining coupling to the conduction electron bands causes the Kondo screening of this spin 1/2 ground state leading a very narrow Kondo peak in the preformed gap structure at odd cluster sites.

\subsubsection{Discussion: Metallic surface conductor in an Kondo insulator}
\label{sec:metallic-surface}

\begin{figure}[t]
\begin{center}
\includegraphics[width=0.482\textwidth,clip]{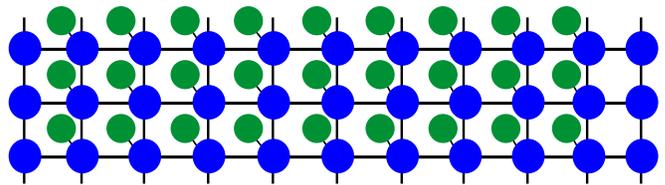}

\caption{Sketch of a strip of a 2d MIAM model for a Kondo insulator with a infinite large y-axis
but a finite size x-axis to define a surface. The correlated orbitals are depicted in green, the uncorrelated in blue. Note that on the surface the correlated orbitals are absent. }
\label{fig:8}
\end{center}
\end{figure}

So far we exemplified a metallic surface $f$-spectra in a small 
MIAM cluster embedded in an metallic environment that is in agreement with
auxiliary field QMC data \cite{RaczkowskiAssaad2019} as well as with the extended Lieb-Mattis theorem.
Since the spatial dimension does not enter in the extended Lieb-Mattis theorem, only requiring 
half-filling, antiferromagnetic coupling and a bi-partite lattice, we can conclude that even the surface of a macroscopically large cluster in 3d will be a surface conductor as long as the number of host lattice sites $N_c\gg N_f$.

On the other hand, we can use the cluster Lieb-Mattis theorem  \cite{Shen1996} to analyze the surface states of
a very large but finite size cluster where $N_f=N_c= O(10^{23})$. Since the removal of a single orbital, a correlated or uncorrelated does not matter, always induces a local moment of $s=1/2$, the spectra must always have a gap independent
of the boundary conditions.
This well known observation inspired many theoretical papers \cite{ColemanTopKI2010,Alexandrov2013,WernerAssaad2013,PetersKamakami2016}  
in which the Kondo lattice model is extended by spin-orbit interactions 
to a topological Kondo insulator in order to address the experimental findings of a 
metallic surface conductance on Kondo insulators \cite{Kim2014,Park2016,SmB6trivalSurfaceConductor2018}.
All these concepts use a perfect lattice structure that abruptly ends at some boundary which appears to be 
very unrealistic in real strongly correlated materials that often have changing surfaces after cleaving. Chemical reactions with oxygen or surface roughness leads to deviation of the surface for an ideal situation. This well known fact made topological
models so popular since a metallic surface state is topologically protected independent of the surface
ruggedness of the real material.

This discrepancy between the ideal trivial Kondo lattice with its insulating surface and a generically
metallic surface in a large Kondo lattice cluster model embedded in a infinitely large metallic host 
raises the question of how many additional uncorrelated orbitals we need to add in order to see a metallic surface
state developing at the edge of the Kondo lattice.

For simplicity let us consider a 2d strip of a Kondo insulator symbolized in Fig.\ \ref{fig:8}. We imagine a very large system in y-direction with $N_y\to \infty$. The number of sites $N_x$  in x-directions remains finite to establish two 
1d surfaces at both edges of the strip. The exact number $N_x$ does not enter the argument and can be arbitrarily large
but only need to be finite in order to establish an edge. The major ingredient in the setup of Fig.\ \ref{fig:8} is the removal of the correlated orbital at the surface layer of a Kondo insulator.
In essence, Fig.\ \ref{fig:8} describes the smallest extension of a pristine finite size Kondo lattice model to a MIAM
by adding a single uncorrelated orbital.

\begin{figure}[t]
\begin{center}
\includegraphics[width=0.5\textwidth]{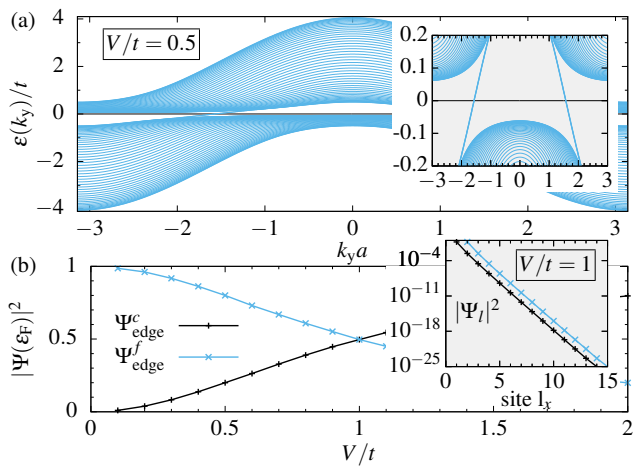}

\caption{(a) The $k_y$-dependent 
band structure for the periodic stripe in Hartree mean field theory, $\tilde \e^f=\e^f+U/2=0$ 
of a Kondo insulator model depicted in 
Fig.\ \ref{fig:8} and  $V_l/t= 0.5$ for a stripe width of $N_x=50$. The inset on the left hand side
reveals the small hybridization gap and the insulating behavior. Clearly visible are the two metallic surface bands that are not crossing each other, one for positive and one for negative $k_y$.
(b) Probability of the edge $f$ and $c$ orbital contribution at the Fermi-energy to the metallic band as function of $V/t$. The inset depicts this probability contribution for $V/t=1$ as function of the site along the x-axis on a logarithmic scale.
}
\label{fig:9}
\end{center}
\end{figure}

In case of a finite $N_y$ the conventional Lieb-Mattis theorem needs to be applied. In analogy to the conventional Kondo insulator, removing a single orbital always leads to a change of the ground state multiplet and, hence, starting from $S^{\rm tot}_z=0$ results in $S^{\rm tot}_z=1/2$, indicating $\rho^{f/c}_l(\w=0)=0$.
In order to clarify the existence of a gapped $f$-spectra at the boundary in the continuum limit $N_y\to \infty$, we invoke the arguments of the extended Lieb-Mattis theorem as outlined before.
In a first step we neglect the two edge stripes of uncorrelated orbitals such that the model results in the conventional Kondo insulator, and a $S^{\rm tot}_z=1/2$ ground state emerges after removing one of the $f$-orbitals at an arbitrary site (not only at the boundary).
However, after including the uncorrelated edge stripes the 
extended Lieb-Mattis theorem predicts a reduction of the moment to $S^{\rm tot}_z=0$ which proves an effective antiferromagnetic coupling of the induced local moment and the 1d edge strip when recovering the original model 
as depicted in Fig.\ \ref{fig:8}: The ground state remains in the SC FP from which we conclude that
all the $f$-spectra must be metallic, $\rho^f_l(\w=0)>0$.  

Since we expect an exponential suppression of the metallic states in the bulk, that is beyond the extended Lieb-Mattis theorem which only distinguishes between $\rho^f_l(\w=0)>0$ and $\rho^f_l(\w=0)=0$, this strongly indicates that even for an infinitely large system adding a layer of uncorrelated orbitals immediately yields a metallic surface even in a conventional Kondo insulator.

Here the metallic behavior is driven by the Kondo effect, while the gap-formation in the Kondo insulator is more related to a self-screening mechanism driven by the RKKY interaction than to a picture of a single ion Kondo effect.

This leaves the question of the role of the correlations and in more details the properties
of the uncorrelated surface orbital defining the two vertical surface boundaries in Fig.\ \ref{fig:8}.
By making use of the adiabatic connection of the ground state properties upon varying the parameter strength
we can answer some aspects by sending the local hybridization $V_l\to \infty$. Then the correlated site form individual bound states with the local conduction orbital comparable with an infinitely strong Kondo coupling.
Consequently, two uncorrelated 1d surface tight binding models emerge which show trivial metallic behavior
determined by the tight-binding model of the edge states. 
This metallic behavior will be preserved for finite $V_l$ and we end up with a trivial metallic surface for arbitrary $V_l$ but
with an unclear mixture of the different orbitals.

We tested this prediction by considering the mean-field theory, i.\ e.\ setting $\tilde \e^f=\e^f+U/2$ which is 
equivalent to $U=0$ while maintaining particle-hole symmetry. 
We solved the single-particle tight binding model in the geometry shown in  Fig.\ \ref{fig:8}
for an infinitely large stripe of width $N_x=50$ with periodic boundary condition in $y$-directions.
The $2N_y-2$ bands are depicted in Fig.\ \ref{fig:9}(a) revealing an insulating behavior with
a finite band gap.  $N_y-2$ bands lay completely below the chemical potential $\mu=0$ and $N_y-2$ bands are located above it.

In addition, two non-crossing metallic bands emerge as shown in the inset of Fig.\ \ref{fig:9}(a).
In order to identify the spatial location of the metallic band and the orbital content, we plot the probability 
contribution of the $c$-edge orbital at $l_x=1$ or $l_x=N_x$ and the $f$-edge orbital
at $l_x=2$ or $l_x=N_x-2$ in Fig.\ \ref{fig:9}(b) as function of the hybridization strength $V$
at the chemical potential. The inset depicts the probability contribution for $V/t=1$ as function of $l_x$ on a logarithmic scale. 
The data reveals that the metallic state is exponentially localized at the edge and $V/t$ determines the mixture
of $c$ and $f$ orbitals. For a small hybridization, the wide band limit, the $f$-orbital content dominates
while with increasing $V$ the uncorrelated edge orbital increases its contribution. For $V=t$, both orbital contribute equally.
The figure also confirms the prediction made above that for infinitely large $V$, only the uncorrelated edge orbitals contribute to the metallic surface band close to the chemical potential. We conjecture that this picture prevails in a model that 
also includes the fluctuating contributions of the Coulomb interaction.

In addition the inset in Fig.\ \ref{fig:9}(b) also reveals that the local density of states
$\rho_{l_x,l_y}^f(0)$ and $\rho_{l_x,l_y}^c(0)$ remain finite in agreement with the finite $U$ extended Lieb-Mattis theorem.
Our $U=0$ analysis and the exponential localization of the metallic surface state in the bulk also reconciles  the seemingly paradox of the prediction of an finite spectra at $\w=0$ by the extended Lieb-Mattis theorem 
and the expectation of a bulk insulator.
The contribution of the metallic surface band to the bulk spectrum at $\w=0$ becomes exponentially small, and for practical purposes
one can not distinguish the bulk spectra from a true insulator with vanishing spectra at $\w=0$.
Since the results for
the $U=0$ spectra is in line with the finite $U$ prediction, there is no QPT at $U=0$ in this case which we attribute to geometric differences between 
infinite and semi-infinite 2d systems. The supercell analysis in Ref.\ \cite{EickhoffKondoHole2021} is valid in arbitrary dimensions, but relies on periodic boundary conditions in all spatial directions.

\subsection{Real-space spectral functions for the Kondo-hole problem}
\label{sec:Kondo-hole}

\begin{figure}[t]
\begin{center}

\includegraphics[width=0.47\textwidth]{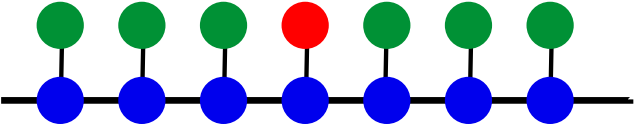}

\caption{Sketch of the geometric setup with a Kondo hole: An array of $N_f = 7$ correlated f-orbitals (green)  are locally 
coupled to a 1D tight binding chain (blue). 
We replace the center correlated orbital with one uncorrelated orbital (red) and introduce $\e^h$ as its single orbital energy.
}

\label{fig:10}
\end{center}
\end{figure}

In this section we address the question how a  Kondo-hole located at the center of a $N_f=7$ MIAM cluster 
changes the spectral properties. The geometry of the problem is sketched in Fig.\ \ref{fig:10}. We 
replace the center 
$f$-orbital by an uncorrelated orbital with an single-particle energy $\e^f_{l_c}=\e^h$ and $U_{l_c}=0$. This allows for modeling  charge neutral substitution of Ce by La
which has a finite $4f$-orbital energy $\e^h\ge 0$ up to complete removing the site by setting $\e^h\to \infty$
\cite{EickhoffKondoHole2021}.

\begin{figure}[t]
\begin{center}
\includegraphics[width=0.5\textwidth,clip]{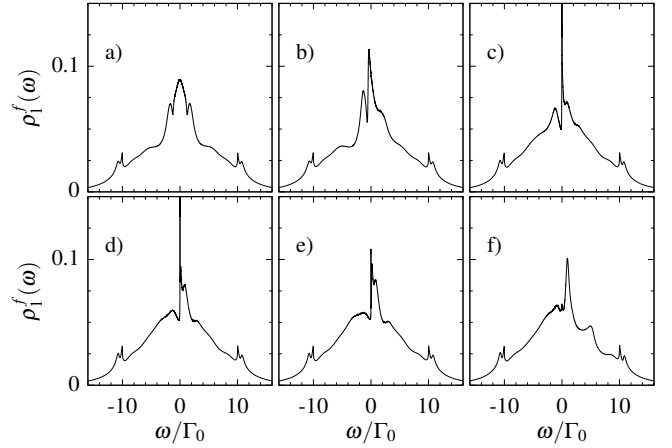}
\caption{Spectral function of the outermost $f$-orbital in an $N_f=7$ impurity array with an uncorrelated hole-orbital in its center at $\e_c=0$. For the correlated $f$-orbitals $\epsilon_f=-U/2=5\Gamma_0$ holds, and the different panels show different onsite-energies of the hole-orbital: (a) $\epsilon^h/\Gamma_0=0.0$, (b) $\epsilon^h/\Gamma_0=1.0$, (c) $\epsilon^h/\Gamma_0=2.0$, (d) $\epsilon^h/\Gamma_0=2.1$, (e) $\epsilon^h/\Gamma_0=2.2$, (f) $\epsilon^h/\Gamma_0=5.0$. The QCP occurs at around $\epsilon^h_c/\Gamma_0\approx 2.25$.}
\label{fig:11}
\end{center}
\end{figure}

Let us start with the evolution of the spectrum at the outermost f-orbital for six different values
$\epsilon^h/\Gamma_0$ plotted in Fig.\ \ref{fig:11}.
The initial cluster  occupation starts out at $N_f=7$ charges
at $\epsilon^h/\Gamma_0=0.0$ and monotonically decreases to $6$ charges for $\e^h\to \infty$ when the central orbital is completely unoccupied and can be removed. 
Therefore, we expect that spectral weight is shifted from $\w<0$ to $\w>0$ upon
increasing $\epsilon^h$. The main shift of the spectral weight is expected to occur for the central orbital.

The system undergoes a quantum phase transition (QPT) \cite{EickhoffKondoHole2021} at critical
value  $\epsilon^h_c/\Gamma_0\approx 2.25$ from a SC FP for $|\epsilon^h|<\epsilon^h_c$ to a local moment FP for $|\epsilon^h|>\epsilon^h_c$in agreement with the extended Lieb-Mattis theorem. The initial PH symmetric spectrum becomes PH asymmetric for 
$\epsilon^h>0$. The screening temperature $T_0$ is exponentially suppressed at $\epsilon^h_c$, typical for a
Kosterlitz-Thouless type transition, and the very narrow Kondo peak  developing at $\w=0$ becomes numerically not resolvable
when approaching the QPT, i.\ e.\  $\epsilon^h\to [\epsilon^h_c]^-$. However, the boundary or surface spectrum remains metallic at any time.

\begin{figure}[t]
\begin{center}
\includegraphics[width=0.49\textwidth,clip]{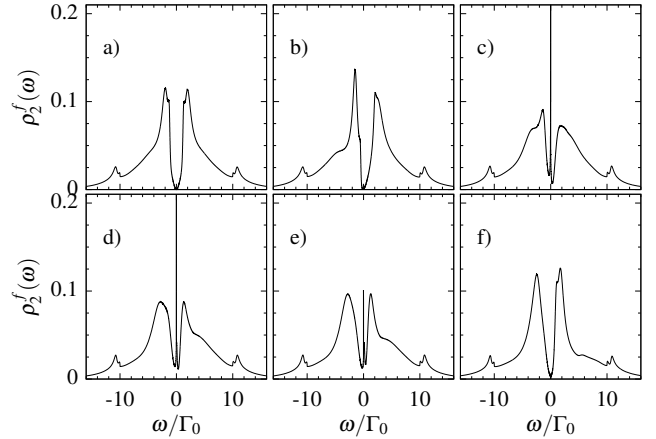}
\caption{Same as Fig. \ref{fig:11} but for the second $f$-orbital.}
\label{fig:12}
\end{center}
\end{figure}

The situation changes when inspecting the spectral evolution for the second correlated orbital $l=2$  
as depicted in Fig.\ \ref{fig:12}
for the same parameters as in the previous figure. 
As in Fig.\ \ref{fig:4}(b), the spectra vanishes at $\w=0$ for 
$\epsilon^h=0$
\footnote{Note  that $U_{l_c}=0$ in this case in contrary to Fig.\ \ref{fig:4}. Therefore,
 the $\epsilon^h=0$ spectra at different sites are not fully identical to those plotted in Fig.\ \ref{fig:4}.}.
Also the extended Lieb-Mattis theorem is not applicable for finite $\epsilon^h$ since the system is shifted away from half-filling. 
A quasi-bound state is developing at the two adjacent sites to the Kondo hole, $l=3$ and $l=5$, whose local moment is screened
via a coupling mediated by the central uncorrelated Kondo hole orbital. Once the low-energy (Kondo) scale 
is  small enough, a very narrow Kondo peak develops inside of the gap as can be seen in Fig.\ \ref{fig:12}(c).
Furthermore, the pseudo-gap is absent ,and $\rho^f_{2,\sigma}(0)>0$. This additional Kondo effect well below the coherent
temperature of the $N_f=7$ cluster destroys the Kondo insulator properties in the spectra. The system approaches 
the KT type QPT when increasing $\epsilon^h\to \epsilon^h_c$. The low temperature screening scale is already exponentially suppressed in Fig.\ \ref{fig:12}(e) such that the peak is very hard to resolve. In the local moment phase, 
depicted in  Fig.\ \ref{fig:12}(f) the
Kondo peak is absent in the spectrum leaving us with a slightly asymmetric spectrum with an onset of a pseudo-gap.
For $\epsilon^h\to\infty$, we recover a half-filled system with one correlated site removed. The extended Lieb-Mattis theorem predicts
a true pseudo-gap spectrum with $\rho_{2,\sigma}^f(0)=0$,
consistent with the spectrum in the local moment phase as depicted in 
Fig.\ \ref{fig:12}(f).
Note, however, that  $\rho_{2,\sigma}^f(0)$ is very small at finite $\epsilon^h$ 
and only vanishes in the limit $\epsilon^h\to\infty$ where the extended Lieb-Mattis theorem is applicable.

\begin{figure}[t]
\begin{center}
\includegraphics[width=0.49\textwidth,clip]{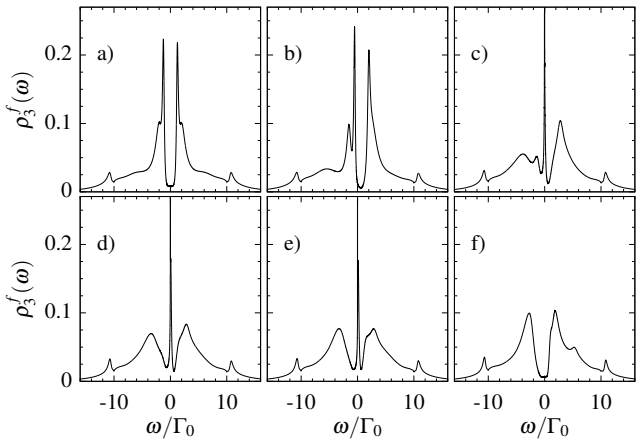}
\caption{Evolution of $\rho_{3,\sigma}^f(\w)$  neighboring the hole site for the same
parameters as in Fig. \ref{fig:11}.}
\label{fig:13}
\end{center}
\end{figure}

Figure \ref{fig:13} shows the evolution of spectra for the third side using the same parameters as in the previous two figures. On top of a spectral valley around $\w=0$ with $\rho_{3,\sigma}^f(0)>0$, a Kondo peak emerges 
at $\w=0$ due to the screening of local moment that is created by the Kondo hole
as long as $\epsilon^h<\epsilon^h_c$.
This Kondo peak also disappears at the QPT once we enter the local moment regime. 
However, an additional finite frequency peak is clearly visible at finite but small $\w$
in the spectral function plotted in Fig.\ \ref{fig:13}(f) whose origin we discuss below.

\begin{figure}[t]
\begin{center}
\includegraphics[width=0.49\textwidth,clip]{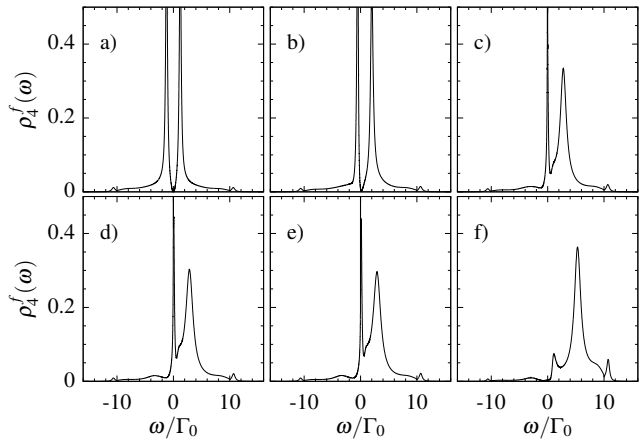}
\caption{Evolution of the Kondo hole spectrum $\rho_{4,\sigma}^f(\w) $ in the cluster center
for the same parameters as in Fig. \ref{fig:11}.}
\label{fig:14}
\end{center}
\end{figure}

With this basic understanding of spectral evolution we draw our attention to the spectrum at the Kondo hole site
hosting an uncorrelated $f$-orbital whose single-particle energy is increased across the KT-type transition in the panels from Fig.\ \ref{fig:14}(a) to Fig.\ \ref{fig:14}(f) 
 using the same parameters as for the previous three figures.
Starting from a PH symmetric spectrum with  a gap, i.\ e.\ $\rho_{4,\sigma}^f(0)=0$, for $\epsilon^h=0$ 
the spectrum becomes increasing more asymmetric, and the gap disappears. We also observe the substantial 
shift of the spectral weight from below to above the chemical potential reflecting 
that the  Kondo hole occupations is rapidly and monotonically decreasing for $1<\epsilon^h/\Gamma_0\to \infty$. 
In fact, the crossover from half-filling to an empty orbital occurs just before $\e^h_c$ but does not coincide with 
the QPT.

Starting from Fig.\ \ref{fig:14}(c) the main single-particle spectral peak from the 
local orbital is clearly visible at $0<\w\approx \epsilon^h$. 
Simultaneously the spectrum show and additional peak that crosses the Fermi-energy just below the critical value
$\e^h_c$. Its  spectral contribution is small and is overshadowed by a very narrow Kondo resonance
that disappears in the local moment regime depicted in Fig.\ \ref{fig:14}(f).

\begin{figure}[t]
\begin{center}
\includegraphics[width=0.49\textwidth,clip]{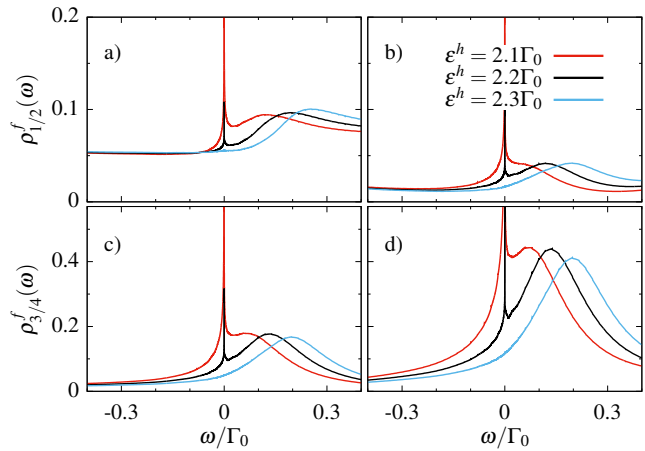}
\caption{Zoom of the spectral functions around the Fermi energy near the QCP located at $\e^h_c/\Gamma_0\approx 2.25$. 
(a) $l=1$, (b) $l=2$, (c) $l=3$, (d) $l=l_c = 4$.
}

\label{fig:15}
\end{center}
\end{figure}

In order to address the spectral evolution across the QPT  in more detail, we 
focus on the low-energy part of the site-dependent spectra for three values $\epsilon^h/\Gamma_0=2.1,2.2,2.3$
and combine the three spectra for each  cluster site into one panel of Fig.\ \ref{fig:15}, starting from
the outermost site $l=1$ in Fig.\ \ref{fig:15}(a) and ending at the center site $l_c=4$ plotted in 
Fig.\ \ref{fig:15}(d). 
We clearly see the Kondo resonances at each site pinned at $\w=0$
which is a result of the screening of the cluster local moment.
The peak height depends on the contribution of the local site to this Kondo effect,
and the resonance is  absent in the LM FP.

In addition, we observe an broader resonance just above the Fermi energy most pronounced
in the spectra around the Kondo hole as plotted in Fig.\ \ref{fig:15}(c)$+$(d). 
In order to clarify its origin, we performed an exact diagonalization of the MIAM Kondo-hole cluster
by just taking into account the inter-site hopping generated by $\mat{T}_\sigma= \Re \mat{\Delta}_\sigma (-i0^+)$ and
ignoring the coupling to the effective band continua  setting $\mat{\Gamma}_\sigma=0$.
It turns out that upon raising of $\e^h$ a low lying excitation in the spectrum is crossing the Fermi-energy and simultaneously the cluster occupation on the $f$-sites decreases from $N=7$ to $N=6$ charges. 
This sharp cluster excitation is broadened by including the hybridization  $\mat{\Gamma}$ to the broader peaks located around $0.1<\w/\Gamma_0<0.3$.
Therefore, this is a contribution from a collective excitation within the effective MIAM cluster.

The QPT, however does not occur exactly at this mixed valence point between $N=6$ and $N=7$: 
It is energetically more favorable to form a SC FP and pay some extra single particle energy $\e^h$
as long as the energy gain in spin screening energy as well as the gain in hybridization energy
is larger than the energy loss by the virtual occupation of the Kondo hole orbital. 
Such behavior is typical for a QPT  changing the ground state degeneracy and
has been discussed in the context of a chemically driven quantum phase transition in a two-molecule Kondo system
\cite{AU-PTCDA-dimer}. This energy argument explains why the orbital peak and the peak of low-lying 
cluster excitation is observed at $\omega >0$ and still a Kondo effect is formed. 

\section{Conclusion}

In this paper we studied the spectral properties of dense 1d MIAM comprising up to $N_f=7$ correlated orbitals using NRG in combination with a recently developed wide band approximation \cite{EickhoffMIAM2020}.
We focused on the parameter regime with a half filled conduction band such that the model describes a Kondo insulator in the limit $N_f\to\infty$ when maintaining PH symmetry.

Increasing $N_f$ a dip in the spectra of the central sites develops rapidly, leading to a full gap already for $N_f=3$. This rapid onset of heavy Fermion formation is in agreement with QMC calculations \cite{RaczkowskiAssaad2019} for a finite size Kondo cluster on a 2d square lattice and is caused by the crossover from a MIAM of the first kind to a MIAM of the second kind \cite{EickhoffMIAM2020}.

In case of $N_f=7$ the gap width in the spectral function of the central orbitals already agrees astonishing well with a full DMFT(NRG) calculation for a Kondo lattice with a Kondo coupling obtained from a Schrieffer-Wolff transformation of the 
corresponding SIAM for $N_f=1$.

Using two techniques, (i) a supercell analysis for the uncorrelated limit and (ii) the extended Lieb-Mattis theorem for the strongly interacting and PH symmetric MIAM on a bi-partite lattice, which originally were developed in order to make a statement on
the nature of the ground state in a Kondo hole problem \cite{EickhoffKondoHole2021}, we predicted the essential feature of the spectra at the Fermi energy of an orbital at an arbitrary site $l$:
If the degeneracy of the ground state is enhanced by removing that orbital the spectra needs to be fully gapped, $\rho_l(\w=0)=0$, whereas it is finite if the degeneracy remains constant, $\rho_l(\w=0)>0$.
These predictions
are in perfect agreement with our 1d NRG calculations and can be summarized as follows:
(i) The spectra of the $f$-orbitals at the edge of the impurity array are always metallic, independent of the strength of correlations.
(ii) In case of nearest neighbor hopping between the $c$-orbitals, the remaining spectra are fully gapped at $\w=0$ in the uncorrelated limit, according to the supercell analysis.
(iii) In the strongly interacting limit either all of the $f$-orbitals acquire a finite spectra at $\w=0$ if the finite size $c$-$f$-orbital cluster is coupled to both of the sublattices of the remaining continuum, or the $\w=0$ spectra oscillates as function of the sublattice index otherwise. Note that our numerical results demonstrate a rapid decrease of $\rho(\w=0)$ with increasing distance to the boundary of the $f$-orbital array such that a fully gapped bulk will be restored in the limit $N_f\to\infty$.

The finding of metallic surface states in the MIAM is in agreement with recent QMC calculations \cite{RaczkowskiAssaad2019} for a finite size Kondo cluster on a 2d square lattice, however, the oscillations of the spectra at $\w=0$ in the bulk have not been reported, even if the Kondo cluster only couples to one sublattice of the remaining continuum.

Since the experimental observation of metallic surface states in Kondo insulators triggered a lot of theoretical research with the focus on topological protection protected mechanisms
caused by spin-orbit coupling, the finding of such states in the MIAM naturally raised the question, of how many additional uncorrelated $c$-orbitals are needed in an extension of the conventional Kondo insulator description to obtain a metallic surface.
Using the Lieb-Mattis theorem in combination with a mean-field calculation we demonstrated that the removal of only the outermost $f$-orbitals in a 2d Kondo insulator is sufficient to realize a metallic band that is exponentially localized at the boundary.

We studied the change in the site dependent spectra upon introducing a Kondo hole in the center of a $N_f=7$ cluster model.
Shifting the onsite energy $\e^h$ of the uncorrelated hole orbital from $\e^h=0$ to $\e^h=\infty$, the model undergoes a Kosterlitz-Thouless type quantum phase transition at a critical $\e^h_c$ which is in accordance with the extended Lieb-Mattis theorem \cite{EickhoffKondoHole2021}.
For $\e^h\to \e^h_c$ a Kondo resonance develops inside the gap of all $f$-orbitals in the cluster, whose width gets exponentially suppressed at the QCP. In addition a collective excitation is observed in form of a sharp resonance that crosses the Fermi energy already for $\e^h<\e^h_c$ and signals the reduction of charge carriers by one due to the removal of the hole orbital.

\begin{acknowledgments}
The authors would like to thank fruitful discussion with Fakher Assaad.
\end{acknowledgments}

%

\end{document}